\documentclass[]{spie}  

\usepackage{amsmath,amsfonts,amssymb}
\usepackage{graphicx}
\usepackage[colorlinks=true, allcolors=blue]{hyperref}

\usepackage{algorithm}
\usepackage{algpseudocode}
\usepackage{float}
\usepackage{subcaption}
\usepackage{booktabs}
\usepackage{graphicx}
\usepackage{indentfirst}
\usepackage{setspace}
\usepackage[skip=3pt]{caption}
\title{Using gradient of Lagrangian function to compute efficient channels for the ideal observer}

\author[a, b]{Weimin Zhou}
\affil[a]{Wyant College of Optical Sciences, University of Arizona, Tucson, USA}
\affil[b]{Department of Medical Imaging, University of Arizona, Tucson, USA}

\authorinfo{Further author information: (Send correspondence to Weimin Zhou)\\Weimin Zhou.: E-mail: weiminzhou@arizona.edu}

\begin{document} 
\maketitle

\begin{abstract}
It is widely accepted that the Bayesian ideal observer (IO) should be used to guide the objective assessment and optimization of medical imaging systems. 
The IO employs complete task-specific information to compute test statistics for making inference decisions and performs optimally in signal detection tasks.
However, the IO test statistic typically depends non-linearly on the image data and cannot be analytically determined. 
The ideal linear observer, known as the Hotelling observer (HO), can sometimes be used as a surrogate for the IO.
However, when image data are high dimensional, HO computation can be difficult.
Efficient channels that can extract task-relevant features have been investigated to reduce the dimensionality of image data to approximate IO and HO performance. 
This work proposes a novel method for generating efficient channels by use of the gradient of a Lagrangian-based loss function that was designed to learn the HO. 
The generated channels are referred to as the Lagrangian-gradient (L-grad) channels.
Numerical studies are conducted that consider binary signal detection tasks involving various backgrounds and signals.
It is demonstrated that channelized HO (CHO) using L-grad channels can produce significantly better signal detection performance compared to the CHO using PLS channels.
Moreover, it is shown that the proposed L-grad method can achieve significantly lower computation time compared to the PLS method.  
\end{abstract}


\section{INTRODUCTION}
\label{sec:intro}  

It has been widely accepted that the Bayesian ideal observer (IO) should be used for objective assessment of medical imaging systems \cite{barrett2013foundations}.
However, the IO test statistic generally is a non-linear function of the image data and cannot be determined analytically.
The Hotelling observer (HO) is the ideal linear observer and can sometimes be utilized as a surrogate for the IO.
However, the computation of the HO can become difficult for high-dimensional image data because a huge covariance matrix needs to be inverted.
Supervised learning-based methods that train neural networks directly to approximate the ideal observer have been actively investigated \cite{kupinski2001ideal, zhou2018learning, zhou2019learning_HO, zhou2019approximating, zhou2020approximating}.
Moreover, Markov-chain Monte Carlo (MCMC)-based methods were developed to estimate IO performance \cite{kupinski2003ideal, zhou2023ideal}.
Another promising approach to addressing the challenges in IO computation is to reduce the dimensionality of the image data by using a set of efficient channels. 

Various efficient channels were developed to extract features from images for approximating the IO and HO for signal detection tasks.
Barrett \emph{et al.} proposed Laguerre-Gauss (LG) channels for detection tasks in which the signal considered is radially symmetric and the correlation in the background has no preferred orientation \cite{barrett1998stabilized}.
Park \emph{et al.} proposed singular vector decomposition (SVD) channels that require knowledge of the system response \cite{park2009singular}.
Witten \emph{et al.} developed partial least squares (PLS) channels that are not limited to specific signal, background, and imaging operator \cite{witten2010partial}.
More recently, Granstedt \emph{et al.} investigated the use of autoencoders to establish efficient channels for approximating the HO \cite{jason2023approximating}.

Zhou \emph{et al.} proposed a Lagrangian-based loss function to learn the HO \cite{zhou2019approximating}.  
Inspired by that work, in this study, I propose a new method for generating efficient channels based on the gradient of the Lagrangian-based loss function that was designed to learn the HO. The channels produced by this method are referred to as Lagrangian-gradient (L-grad) channels.
 Computer simulation studies were conducted that considered signal detection tasks involving a multivariate normal lumpy background (MVNLumpy) and a VICTRE Mammography dataset \cite{badano2018evaluation}.
 It is demonstrated that the proposed L-grad channels can significantly outperform the PLS channels in terms of detection performance and computation~time.

\section{Methods}
This work considers a binary signal detection task that requires an observer to classify image data $\mathbf{g}\in \mathbb{R}^{M\times 1}$ as satisfying the signal-absent hypothesis ($H_0$) or the
signal-present hypothesis ($H_1$). The imaging processes under the two hypotheses are: 
\begin{subequations}
\begin{equation} \label{subeq1} 
H_0: \mathbf{g} = \mathcal{H}\mathbf{f}_b+\mathbf{n} \equiv \mathbf{b} + \mathbf{n}, 
\end{equation}
\begin{equation}
H_1: \mathbf{g} = \mathcal{H}(\mathbf{f}_b+\mathbf{f}_s)+\mathbf{n} \equiv \mathbf{b} + \mathbf{s}+\mathbf{n},
\end{equation}
\end{subequations}
where $\mathcal{H}$ is a continuous-to-discrete (C-D) imaging operator, $\mathbf{f}_b$ and $\mathbf{f}_s$ are background and signal object functions, respectively, $\mathbf{n}$ denotes the measurement noise, and $\mathbf{b}\equiv \mathcal{H}\mathbf{f}_b$ and $\mathbf{s}\equiv \mathcal{H}\mathbf{f}_s$ are the background image and signal image, respectively.
\subsection{Ideal observer and channelized observer}
The Bayesian Ideal observer (IO) employs complete task-specific information in the image data and sets the upper performance limit for binary signal detection tasks.
The IO employs a likelihood ratio as its test statistic: 
\begin{equation}
t_{IO} (\mathbf{g}) = \frac{p(\mathbf{g}| H_1)}{p(\mathbf{g}|H_0)}.
\end{equation}
Generally, $t_{IO}$ is a non-linear function of $\mathbf{g}$ and cannot be analytically determined.
When a non-linear IO test statistic is difficult to compute,
the ideal linear observer, known as the Hotelling observer (HO), can sometimes be employed to assess task-based image quality. The HO test statistic can be calculated as $t_{HO} (\mathbf{g}) = \mathbf{w}_{HO}^T\mathbf{g}$,
where $\mathbf{w}_\text{HO}\in\mathbb{R}^{M\times 1}$ is the Hotelling template~\cite{barrett2013foundations}: 
\begin{equation}
\mathbf{w}_\text{HO}=\left[\frac{1}{2}(\mathbf{K}_0 + \mathbf{K}_1)\right]^{-1}\Delta\bar{\bar{\mathbf{g}}},
\end{equation}
where $\mathbf{K}_j =\big\langle \langle [\mathbf{g} -\bar{\bar{\mathbf{g} }}_j][\mathbf{g} -\bar{\bar{\mathbf{g} }}_j]^{T}  \rangle_{\mathbf{g}|\mathbf{f}} \big\rangle_{\mathbf{f}|H_j}$  is the covariance matrix of the measured image data $\mathbf{g}$ under hypothesis $H_j$ ($j=0, 1$), and $\Delta \bar{\bar{\mathbf{g}}} = \bar{\bar{\mathbf{g} }}_1 - \bar{\bar{\mathbf{g} }}_0$ is the difference between the mean of the image data $\mathbf{g}$ under the two hypotheses. 
Here, $\bar{\bar{\mathbf{g} }}_j\equiv \langle\bar{\mathbf{g}}(\mathbf{f})\rangle_{\mathbf{f}|H_j}$ and $\bar{\mathbf{g}}(\mathbf{f}) = \langle \mathbf{g} \rangle_{\mathbf{g}|\mathbf{f}}$. 
It is sometimes useful to employ covariance matrix decomposition (CMD) for estimating the $\mathbf{K}_j$ for the HO computation:
\begin{equation}\label{eq:K}
\mathbf{K}_j = \big\langle  \langle [\mathbf{g} - \bar{\mathbf{g}}(\mathbf{f})] [\mathbf{g} - \bar{\mathbf{g}}(\mathbf{f})]^T   \rangle_{\mathbf{g}|\mathbf{f}}  \big\rangle_{\mathbf{f}|H_j} 
 + \langle  [\bar{\mathbf{g}}(\mathbf{f}) - \bar{\bar{\mathbf{g}} }_j] [\bar{\mathbf{g}}(\mathbf{f}) - \bar{\bar{\mathbf{g}} }_j]^T  \rangle_{\mathbf{f}|H_j} 
\equiv \langle{\mathbf{K}}_{\mathbf{n}|\mathbf{f}} \rangle_{\mathbf{f}|H_j}+ \mathbf{K}_{\bar{\mathbf{g}}(\mathbf{f}) | H_j}.
\end{equation}
However, direct implementation of the HO has been limited to small-sized images because it requires the inversion of a covariance matrix that can be enormous. 
A popular way to address this limitation is to apply a set of channels that maps high-dimensional image data $\mathbf{g}$ to a low-dimensional vector $\mathbf{v}$: $\mathbf{v} = \mathbf{T}\mathbf{g}$, where $\mathbf{v}$ is a $D\times 1$ vector and $\mathbf{T}$ is a $D\times M$ channel matrix ($D \ll M$). After channelization, the low-dimensional vector $\mathbf{v}$ is subsequently used to perform signal detection tasks. Applying the HO to the channelized data vector $\mathbf{v}$ produces the channelized Hotelling observer (CHO), which computes the test statistic as:
\begin{equation}
    t_{HO}(\mathbf{v}) = \mathbf{w}_v^T\mathbf{v} = (\mathbf{K}_v^{-1}\Delta\bar{\bar{\mathbf{v}}})^T\mathbf{v}
    =  (\mathbf{K}_v^{-1}\Delta\bar{\bar{\mathbf{v}}})^T\mathbf{T} \mathbf{g} \equiv \mathbf{w}_{CHO}^T\mathbf{g}.
\end{equation}
Here, $\mathbf{w}_v$ is the Hotelling template acting on the channelized data vector $\mathbf{v}$, $\mathbf{K}_v = \frac{1}{2}(\mathbf{K}_{v,0} + \mathbf{K}_{v,1})$ is the covariance matrix of the channelized data vector $\mathbf{v}$ averaged between the hypotheses $H_0$ and $H_1$, $\Delta\bar{\bar{\mathbf{v}}} = \mathbf{T}\Delta\bar{\bar{\mathbf{g}}}$ is the difference between the mean of the channelized data vector $\mathbf{v}$ under the two hypotheses, and $\mathbf{w}_{CHO} \equiv \mathbf{T} ^T (\mathbf{K}_{\mathbf{v}}^{-1}\Delta\bar{\bar{\mathbf{v}}})$ is the CHO template acting on the original image data $\mathbf{g}$.

\subsection{Lagrangian-gradient channels}
Zhou \emph{et. al} proposed a supervised learning-based method to approximate the HO by maximizing signal detectability, quantified by the signal-to-noise ratio of test statistics, by solving the following constrained optimization problem \cite{zhou2019approximating}:
\begin{equation} \label{eq:opt}
\begin{split}
 &\underset{\mathbf{w}}{\text{minimize} } 
 \frac{1}{2}\left \langle [\mathbf{w}^T\mathbf{g}-  \mathbf{w}^T\bar{\bar{\mathbf{g}}}_0]^2 \right\rangle_0 + \frac{1}{2}\left \langle [\mathbf{w}^T\mathbf{g}-  \mathbf{w}^T\bar{\bar{\mathbf{g}}}_1]^2 \right\rangle_1, \\
 &\text{subject to\ }  \mathbf{w}^T\bar{\bar{\mathbf{g}}}_1 - \mathbf{w}^T\bar{\bar{\mathbf{g}}}_0=C,
 \end{split}
 \end{equation}
where $C$ is an arbitrary constant that does not affect the optimal solution.
 The Lagrangian function associated with this constrained optimization problem is:
 \begin{equation}\label{eq:L1}
 \begin{split}
 L(\mathbf{w}, \lambda) =  \frac{1}{2}\left \langle [\mathbf{w}^T\mathbf{g}-  \mathbf{w}^T\bar{\bar{\mathbf{g}}}_0]^2 \right\rangle_0 + \frac{1}{2}\left \langle [\mathbf{w}^T\mathbf{g}-  \mathbf{w}^T\bar{\bar{\mathbf{g}}}_1]^2 \right\rangle_1 
 - \lambda ( \mathbf{w}^T\bar{\bar{\mathbf{g}}}_1 - \mathbf{w}^T\bar{\bar{\mathbf{g}}}_0-C),
 \end{split}
 \end{equation}
where $\lambda$ is the Langrage multiplier. The global minimum of $L(\mathbf{w}, 2)$ has been shown to be equal to the Hotelling template $\mathbf{w}_{HO}$ \cite{zhou2019approximating}.

In this work, I propose a new method to produce efficient channels based on the gradient of the Lagrangian function $L(\mathbf{w}, 2)$.
The produced channels are referred to as the Lagrangian-gradient (L-grad) channels.
The gradient of $L(\mathbf{w}, 2)$ can be computed as:
\begin{equation}
\label{eq:grad}
\nabla_{\mathbf{w}}L(\mathbf{w}, 2) =\left[ \mathbf{K}_0 + \mathbf{K}_1\right] \mathbf{w} -2  \Delta \bar{\bar{\mathbf{g}}} . 
\end{equation}

The $-\frac{1}{2}\nabla_{\mathbf{w}}L(\mathbf{w}, 2)$ evaluated at $\mathbf{w}=\mathbf{0}$ corresponds to the non-prewhitening matched filter. 
The L-grad channels are generated by using the following iterative algorithm:
\begin{algorithmic}[1]
\State Estimate $\mathbf{K}_0$, $\mathbf{K}_1$, and $\Delta \bar{\bar{\mathbf{g}}}$ based on the training dataset and/or knowledge of the background and signal.
\State Let $i = 1$ and initialize $\mathbf{w}_1 = \mathbf{0}$.
\State Compute the channel vector $\mathbf{t}_i = -\frac{1}{2}\nabla_{\mathbf{w}}L(\mathbf{w}_i, 2)$, and let its transpose $\mathbf{t}_i^T$ be the $i$th row of the channel matrix $\mathbf{T}_i$. 
\State Evaluate the channelized data $\mathbf{v}_i = \mathbf{T}_i \mathbf{g}$ and compute its covariance matrix $\mathbf{K}_{\mathbf{v}_i}$.
\State 
Compute the CHO template using the first $i$ channel vectors:
$\mathbf{w}_{CHO}^{(i)} = \mathbf{T}_i ^T (\mathbf{K}_{\mathbf{v}_i}^{-1}\Delta\bar{\bar{\mathbf{v}}}_i)$. 
\State $\mathbf{w}_{i+1} \gets \mathbf{w}_{CHO}^{(i)}$ and $i \gets i+1$.
 \State Iterate steps 3--6 to calculate the desired number of channel vectors.
\end{algorithmic}

It should be noted that Step 5 requires an inversion of the covariance matrix of the channelized data.
In the $i$th iteration,
the matrix inversion of $\mathbf{K}_{\mathbf{v}_i}$ can be solved using the matrix inversion lemma given that the covariance matrix $\mathbf{K}_{\mathbf{v}_i}$ can be partitioned into submatrices, including the covariance matrix of the channelized data from all previous $i-1$ iterations $\mathbf{K}_{\mathbf{v}_{i-1}}$, the covariance between channelized data from all previous $i-1$ iterations $\mathbf{v}_{i-1}$ and the new channelized data in the current $i$th iteration $\mathbf{t}_i^T\mathbf{g}$, and the variance of the new channelized data $\mathbf{t}_i^T\mathbf{g}$. In this way, direct inversion of the complete matrix $\mathbf{K}_{\mathbf{v}_i}$ can be avoided, which accelerates the channel generation process.

\section{Numerical studies and results}
\label{sec:sections}

Computer simulation studies were conducted, and two different signal-known-exactly (SKE) detection tasks were considered. In these SKE tasks, the known signal image $\mathbf{s}$ was used as the ground truth $\Delta\bar{\bar{\mathbf{g}}}$ to calculate $\nabla_{\mathbf{w}}L(\mathbf{w}, 2)$ to generate L-grad channels. 
Details about the signal detection tasks considered and the results are shown below.
 
\subsection{SKE detection task with MVNLumpy and Gaussian signal}
\begin{figure}[H]
     \centering
         \centering
         \includegraphics[width=0.9\textwidth]{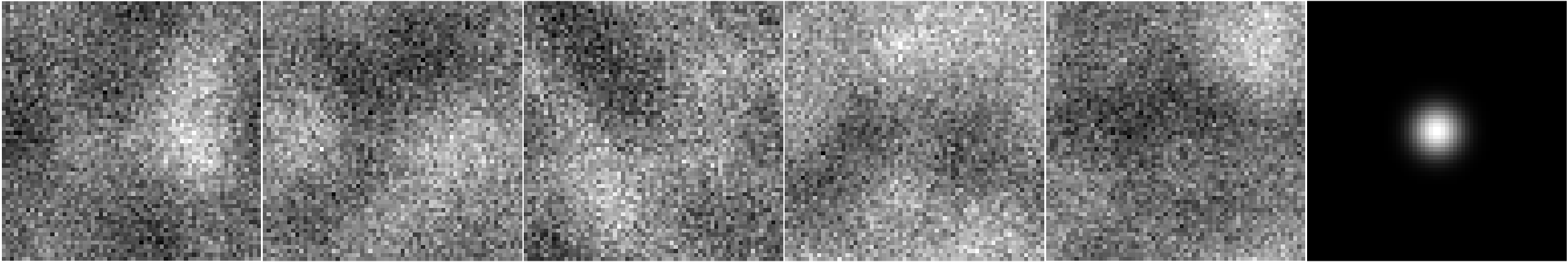}
         \caption{
         The five images from left to right are examples of signal-present images associated with the considered MVNLumpy. Gaussian noise was added to simulate the measured image data. The rightmost image shows the Gaussian signal to be detected.}
         \label{fig:MVN_Lumpy_imgs}
     \end{figure}

A multivariate normal distributed lumpy background (MVNLumpy, also known as type 2 lumpy background) was considered \cite{rolland1990factors}. The to-be-detected signal has a Gaussian profile and the measurement noise follows Gaussian distribution. For this task, the likelihood function under each hypothesis is Gaussian and the HO is equivalent to the IO.
Examples of signal-present images and the signal to be detected are shown in Fig. \ref{fig:MVN_Lumpy_imgs}.

The L-grad channels were produced by use of training datasets with various sizes, including 1000, 2000, 4000, 6000, 8000, 10000, 20000, 30000, and 40000 total images. Half of the images in each training dataset are signal-absent, and the other half are signal-present. 
After the channels were produced, the corresponding CHO was subsequently computed on another set of 4000 images that comprised 2000 signal-absent images and 2000 signal-present images.

The covariance matrix decomposition (CMD) was also used to calculate the L-grad channels, where the covariance matrices $\mathbf{K}_0$ and $\mathbf{K}_1$ were computed using background images and knowledge of the noise statistic. The channels produced in this way are referred to as the L-grad-CMD channels. 
Both L-grad and L-grad-CMD channels were compared to the PLS channels and the reference HO, which was calculated by inverting the covariance matrix estimated using the CMD on 40000 background images.

\begin{figure}[H]
     \centering
     \begin{subfigure}[b]{0.495\textwidth}
         \centering
         \includegraphics[width=\textwidth]{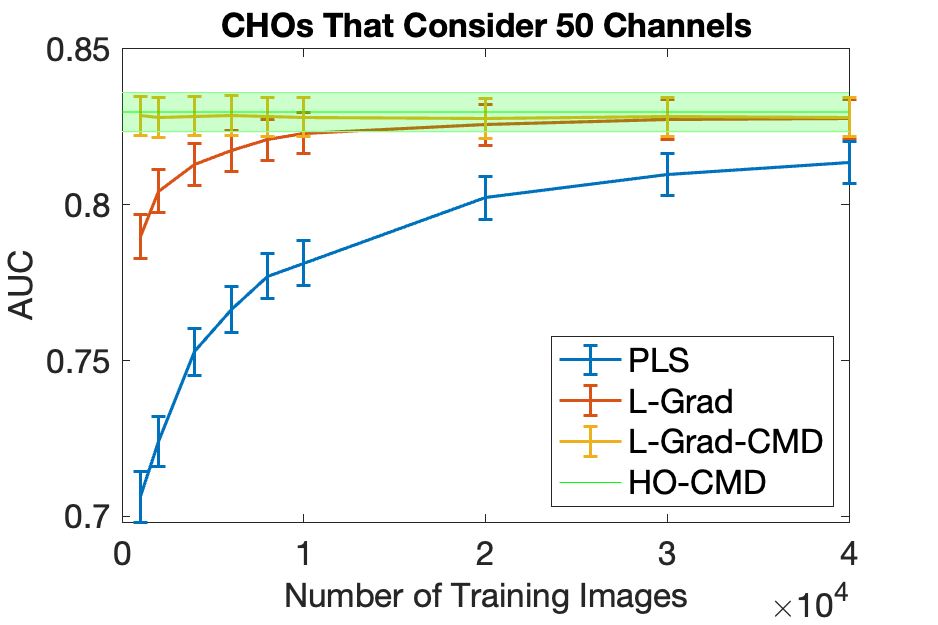}
         \caption{}
     \end{subfigure}
     \hfill
     \begin{subfigure}[b]{0.495\textwidth}
         \centering
         \includegraphics[width=\textwidth]{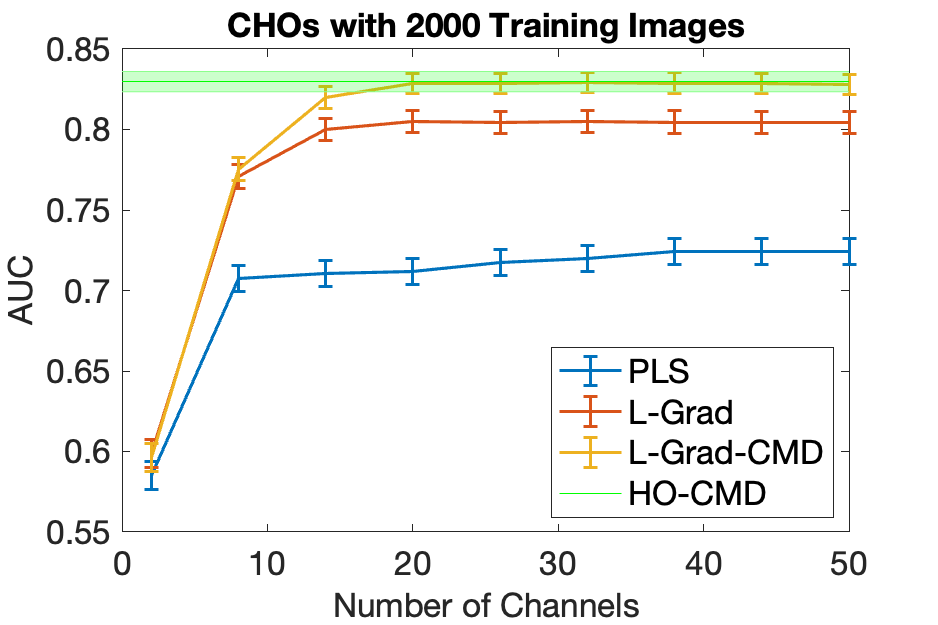}
         \caption{}
     \end{subfigure}
          \caption{(a) CHO performance with 50 L-grad, L-grad-CMD, and PLS channels computed on training dataset with different sizes. (b) CHO performance as a function of the number of channels when 2000 training images were considered.}
          \label{fig:lumpy_curves}
     \end{figure}

The signal detection performance, summarized by the area under the receiver operating characteristic (ROC) curve (AUC) value, was evaluated using a separate testing dataset that comprised 2000 signal-absent images and 2000 signal-present images. The AUC values produced by the CHOs corresponding to the L-grad channels, L-grad-CMD channels, and PLS channels are shown in Fig. \ref{fig:lumpy_curves}.
The curves of AUC values with respect to the number of training images for CHOs with 50 channels are shown in Fig. \ref{fig:lumpy_curves} (a), and the curves of AUC values with respect to the number of channels for CHOs trained with 2000 images are shown in Fig. \ref{fig:lumpy_curves} (b).
The L-grad-CMD channels produced the best AUC values, and both L-grad-CMD and L-grad outperformed the PLS channels.

\begin{figure}[H]
     \centering
     \begin{subfigure}[b]{0.31\textwidth}
         \centering
         \includegraphics[width=\textwidth]{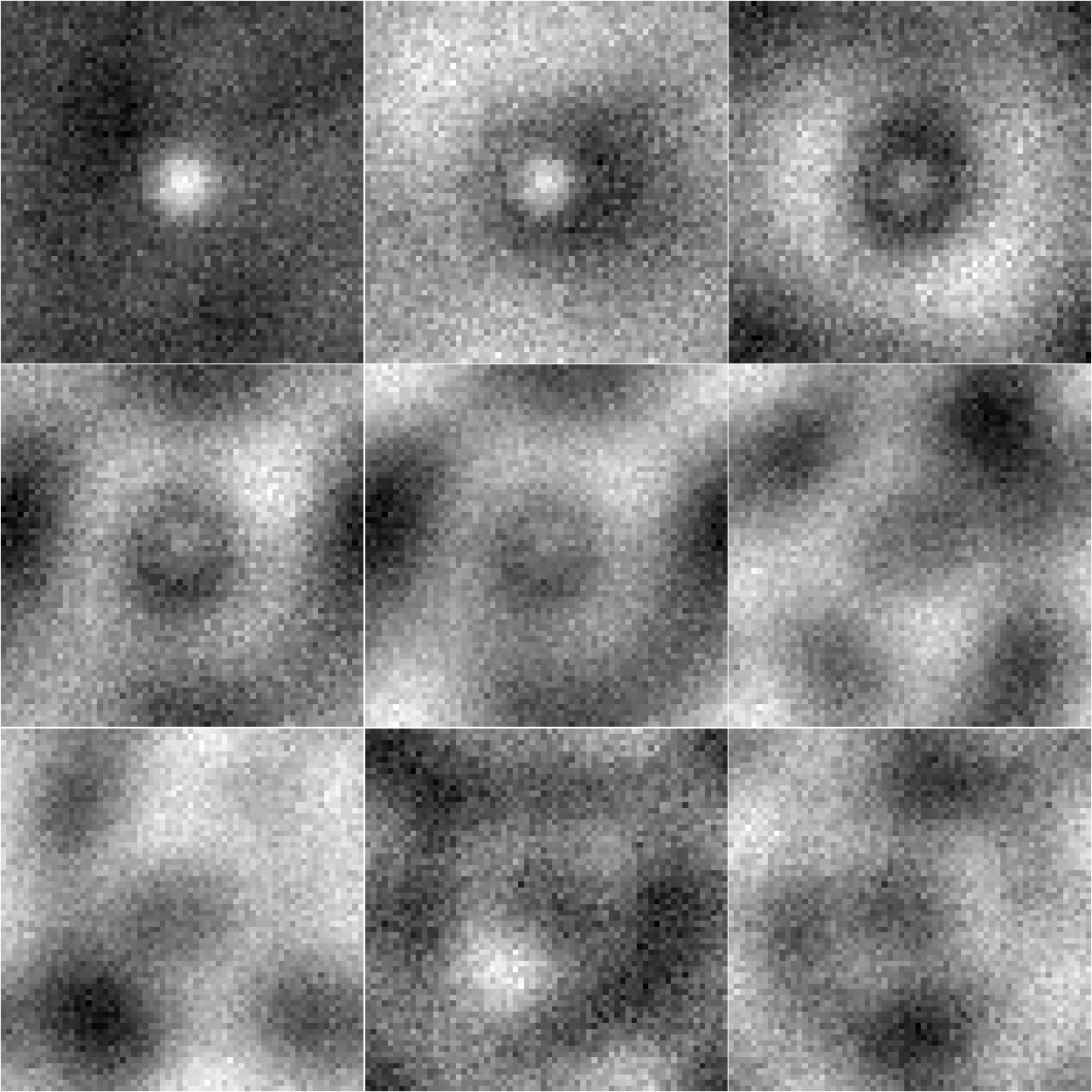}
         \caption{}
     \end{subfigure}
     \hspace{0.1cm}
     \begin{subfigure}[b]{0.31\textwidth}
         \centering
         \includegraphics[width=\textwidth]{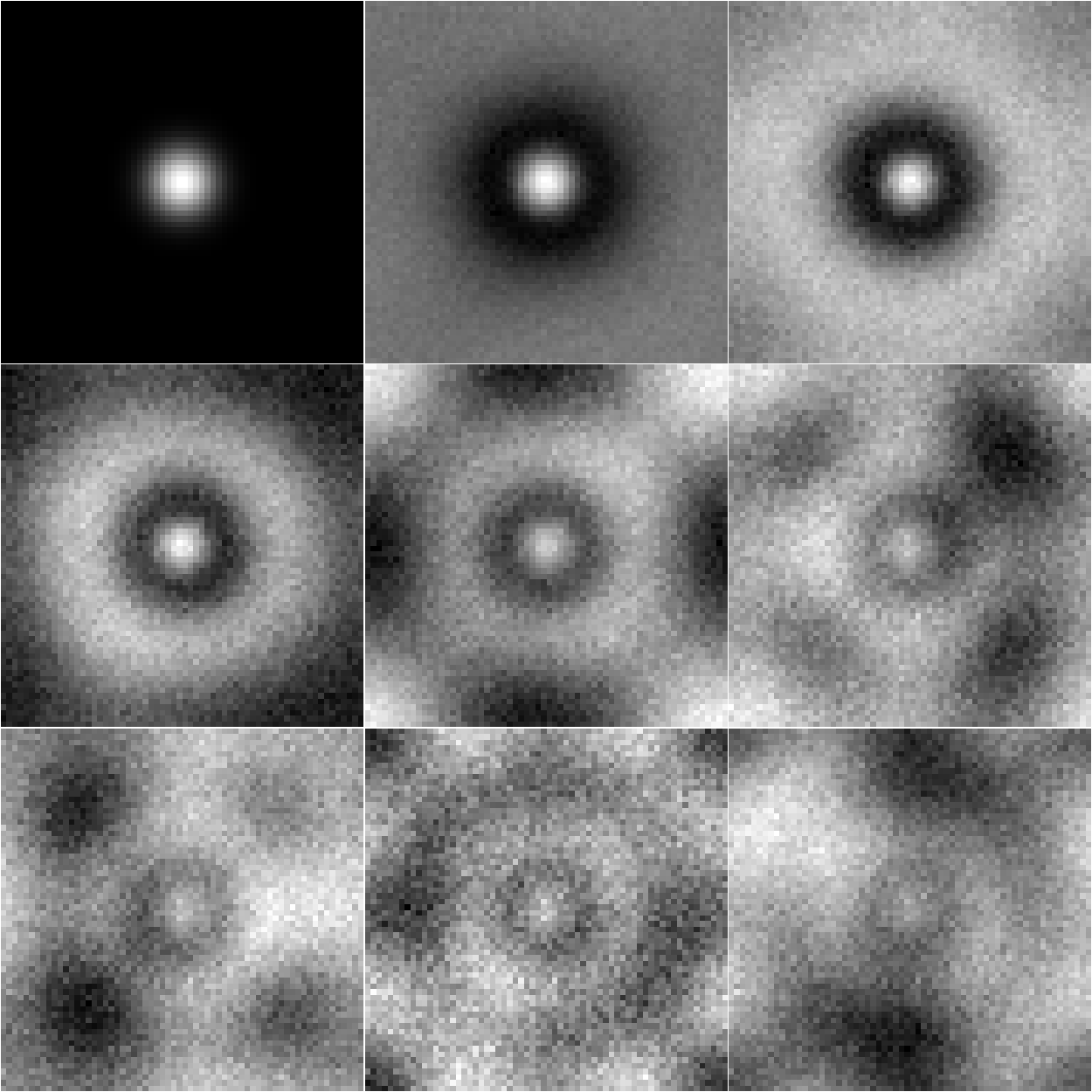}
         \caption{}
     \end{subfigure}
          \hspace{0.1cm}
     \begin{subfigure}[b]{0.31\textwidth}
         \centering
         \includegraphics[width=\textwidth]{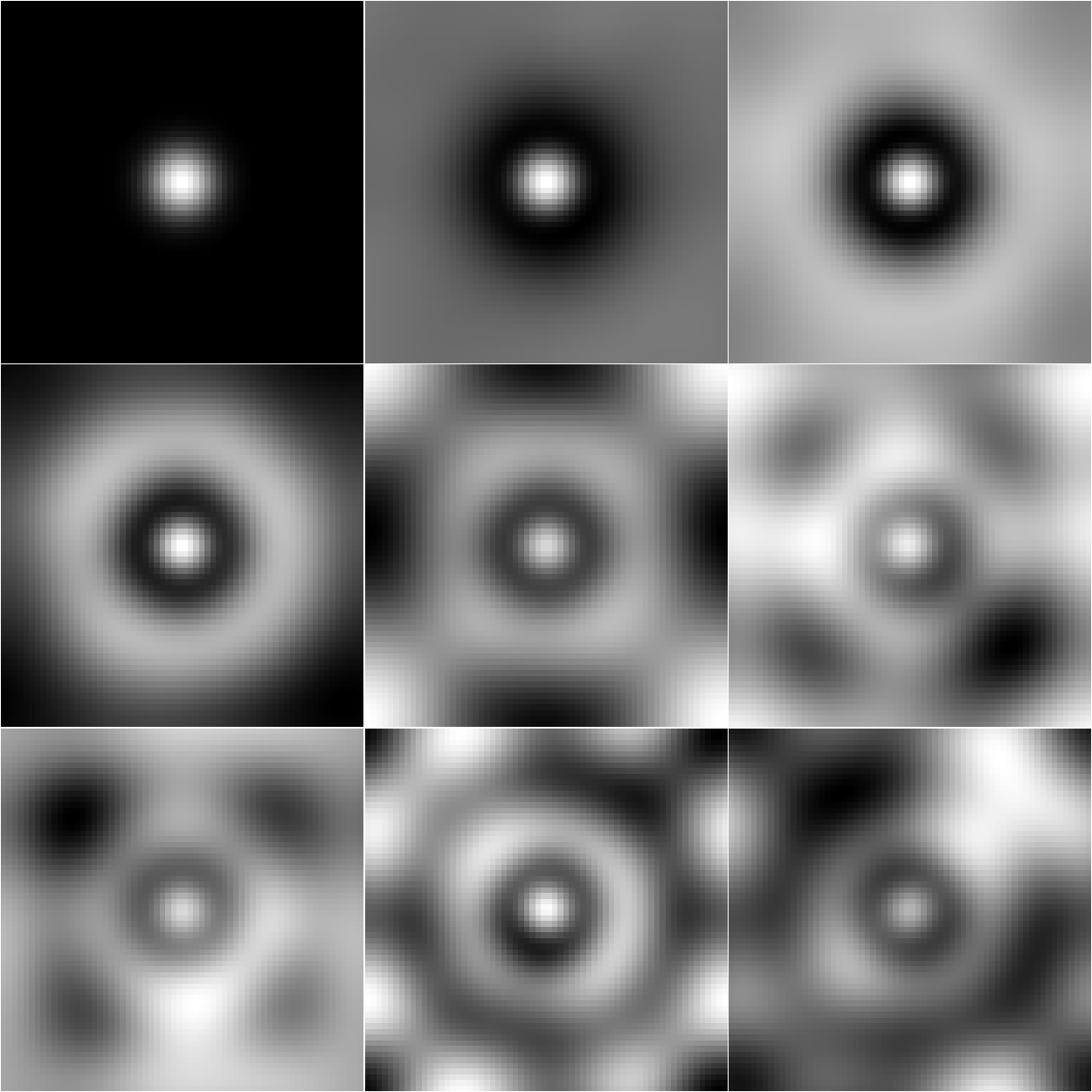}
         \caption{}
     \end{subfigure}
     \caption{The first nine channels produced by PLS (a), L-grad (b), and L-grad-CMD (c).}
     \label{fig:mvn_channels}
     \end{figure}

The first nine channels produced by the PLS, L-grad and L-grad-CMD that employed 2000 training images are shown in Fig. \ref{fig:mvn_channels}. 
The PLS channels, which were generated on noisy training images, are most noisy, while the L-grad channels that incorporated the known signal image are less noisy, and the L-grad-CMD channels that employed both the known signal image and the background statistic are clean.

\vspace{-0.2cm}
The computation time (averaged over 10 trials using MATLAB R2024b on an Apple M4 Max chip) for generating 50 L-grad channels is compared to that for generating 50 PLS channels in Table. \ref{tab1}. The L-grad channels were generated much faster than the PLS channels for all the considered training data sizes. 
In the case of 40000 training images, the computation of L-grad channels was nearly seven times faster than that of the PLS channels.
This is because the computation time of the PLS method is significantly affected by the number of training images, as the PLS needs to employ the entire data matrix in every iteration to generate the channels.
\begin{table}[H]
\centering
\caption{Computation time (in seconds) for generating PLS channels and L-grad channels when different numbers of training images are employed. }
\begin{tabular}{@{}llllllllll@{}}
\toprule
Number of training images & 1000   & 2000     & 6000     & 10000  & 20000  & 30000   & 40000   \\ \midrule
PLS computation time          & 0.5610 & 0.8513  & 1.6240  & 2.0403 & 8.0119 & 11.1468 & 13.7784 \\
L-grad computation time       & 0.1722 & 0.2092  & 0.3637  & 0.4900 & 0.9078 & 1.3251  & 1.7663  \\ \bottomrule
\end{tabular}
    \label{tab1}
\end{table}

\subsection{SKE detection task with VICTRE Mammography ROIs and spiculated mass}

\begin{figure}[H]
     \centering
         \includegraphics[width=0.9\textwidth]{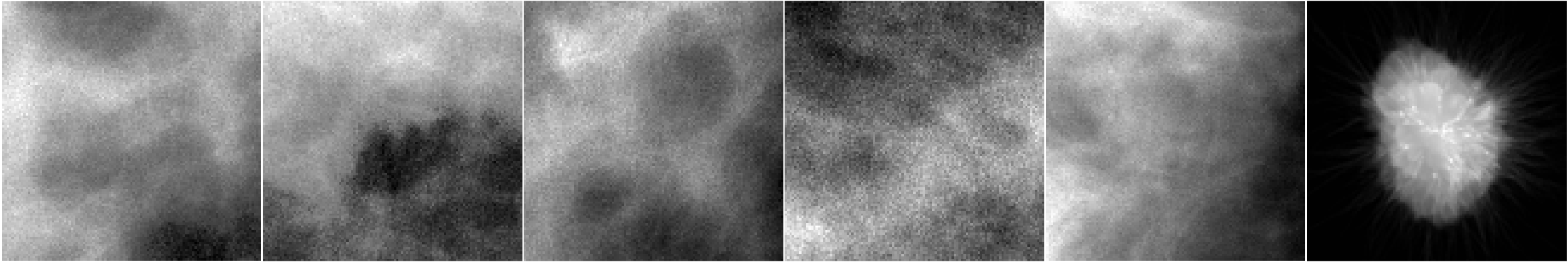}
         \caption{From left to right: Five examples of signal-present images generated by use of VICTRE mammography ROIs and the spiculated mass signal to be detected.}
         \label{fig:VIC_imgs}
     \end{figure}

A set of signal-absent VICTRE Mammography regions of interest (ROIs) (\url{https://github.com/DIDSR/VICTRE_DM_ROIs}) corresponding to four breast types, including extremely dense, fatty, heterogeneously dense, and scattered density, were employed.
The dimension of those ROIs is $109\times 109$.
A 3D spiculated mass signal from a VICTRE lesion model (\url{https://github.com/DIDSR/VICTRE/tree/master/Lesion%20Insertion/VICTRE_LesionModels}) was projected onto a 2D signal image, which was resized to the dimension of $109\times 109$ and subsequently inserted into half of the ROIs to form a set of signal-present images. 
Examples of VICTRE Mammography ROIs and the spiculated mass signal are shown in Fig. \ref{fig:VIC_imgs}.

\begin{figure}[H]
     \centering
     \begin{subfigure}[b]{0.495\textwidth}
         \centering
         \includegraphics[width=\textwidth]{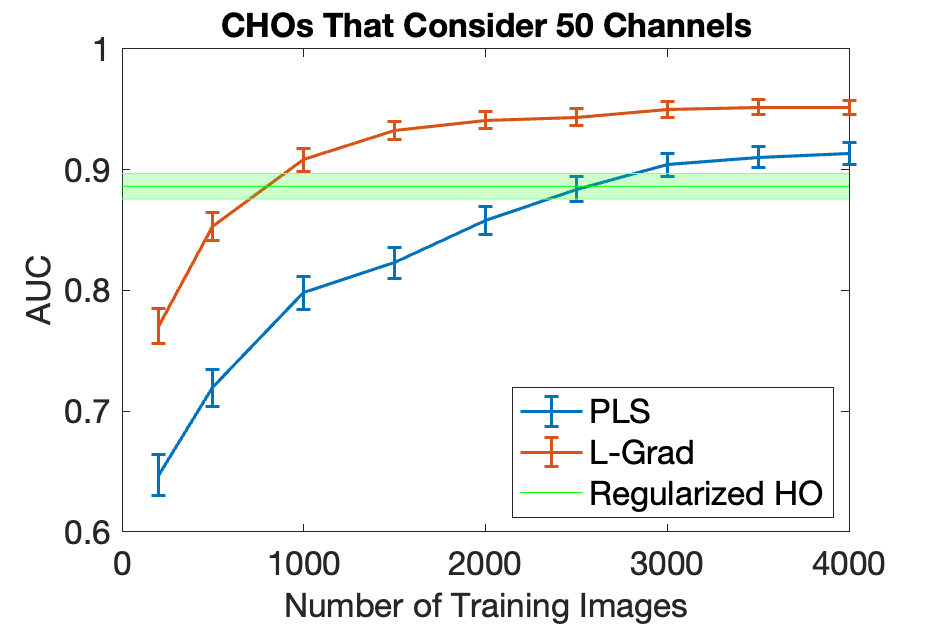}
         \caption{}
     \end{subfigure}
     \hfill
     \begin{subfigure}[b]{0.495\textwidth}
         \centering
         \includegraphics[width=\textwidth]{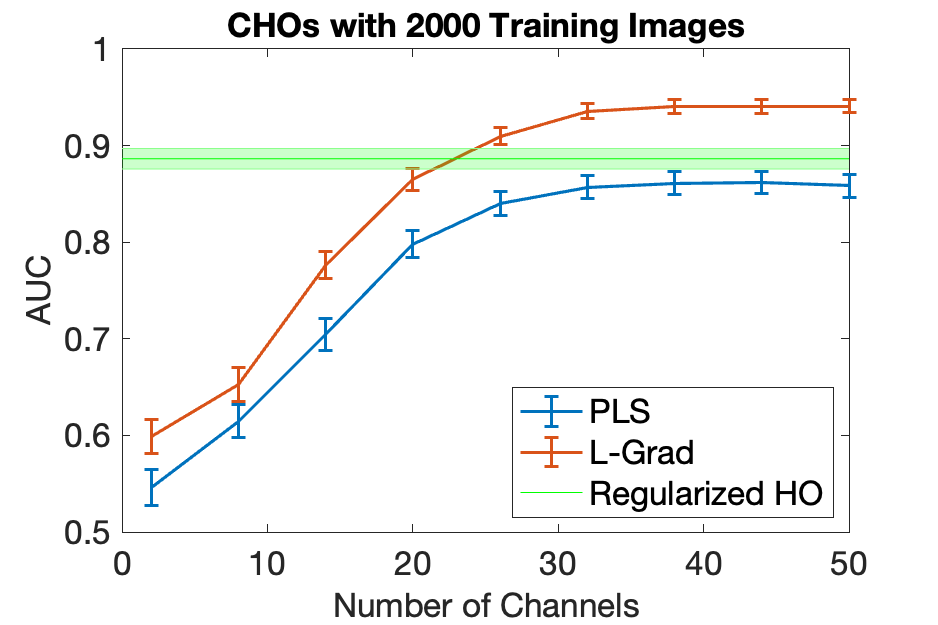}
         \caption{}
     \end{subfigure}
            \caption{(a) CHO performance with L-grad and PLS channels computed with varying numbers of training images for the VICTRE ROIs. (b) CHO performance as a function of the number of channels when the channels were trained on 2000 ROIs.}
            \label{fig:vic_curves}
     \end{figure}

The L-grad channels were produced by use of training datasets with different sizes, including 200, 500, 1000, 1500, 2000, 2500, 3000, 3500, 4000 images. 
Half of the training ROIs are
signal-absent, and the other half are signal-present.
After the channels were produced,
the CHO was computed by use of another 1000 images (500 signal-absent images and 500 signal-present images), and the AUC values were evaluated on a separate set of 1000 testing images (500 signal-absent images and 500 signal-present images). The AUC values corresponding to the L-grad channels and the PLS channels are shown in Fig. \ref{fig:vic_curves}. The L-grad channels produced greater AUC values than the PLS
channels when varying numbers of training images and channels were considered.

Because only a total of 4000 training images were considered and the ROI dimension is $109\times 109$, the covariance matrix of image data is ill-conditioned and cannot be stably inverted to provide an HO study. As an alternative, a regularized HO (RHO) \cite{li2021assessing} was calculated that employs the Moore-Penrose pseudoinverse of the covariance matrix of 4000 ROIs. The RHO performance is compared to the CHO performance in Fig. \ref{fig:vic_curves}.
The L-grad method can achieve higher AUC values than the RHO when the number of channels exceeded 20 and only 2000 training images were used. 

\begin{figure}[H]
     \centering
         \centering
         \includegraphics[width=\textwidth]{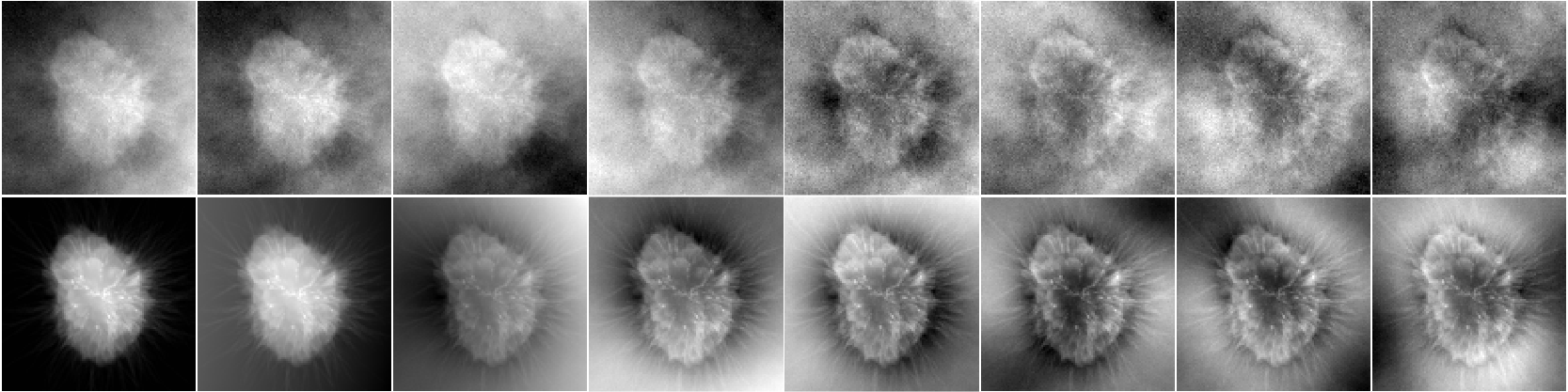}
         \caption{The first eight PLS channels (top) and L-grad channels (bottom) produced by use of 2000 images. The L-grad channels appear cleaner.}
         \label{fig:vic_channels}
     \end{figure}

The first eight channels corresponding to the PLS and L-grad are shown in Fig. \ref{fig:vic_channels}. The channels were produced using 2000 images. 
As expected, the L-grad channels appear cleaner than the PLS-channels as the ground truth signal image was employed in the process of generating the L-grad channels.

\section{Conclusion}
In this work, a novel method for generating efficient channels, which are referred to as the L-grad channels, was introduced. The L-grad channels are produced based on the gradient of a Lagrangian function that was designed to learn the Hotelling observer. Knowledge of the signal and background statistics can readily be incorporated in the proposed L-grad method to produce the efficient channels. The ability of L-grad channels to approximate the
ideal linear observer was investigated in two different signal detection tasks in which an MVNLumpy and a VICTRE
mammography dataset were considered. It has been demonstrated that the L-grad channels can significantly outperform the PLS channels to approximate the ideal linear observer, and the L-grad method can be implemented much faster than the PLS method.

\bibliography{report} 
\bibliographystyle{spiebib} 

\end{document}